# Towards Fast Region Adaptive Ultrasound Beamformer for Plane Wave Imaging Using Convolutional Neural Networks*

Roshan P Mathews, Student *Member,* and Mahesh Raveendranatha Panicker, Senior *Member, IEEE*

*Abstract*— Automatic learning algorithms for improving the image quality of diagnostic B-mode ultrasound (US) images have been gaining popularity in the recent past. In this work, a novel convolutional neural network (CNN) is trained using time of flight corrected in-vivo receiver data of plane wave transmit to produce corresponding high-quality minimum variance distortion less response (MVDR) beamformed image. A comprehensive performance comparison in terms of qualitative and quantitative measures for fully connected neural network (FCNN), the proposed CNN architecture, MVDR and Delay and Sum (DAS) using the dataset from Plane wave Imaging Challenge in Ultrasound (PICMUS) is also reported in this work. The CNN architecture can leverage the spatial information and will be more region adaptive during the beamforming process. This is evident from the improvement seen over the baseline FCNN approach and conventional MVDR beamformer, both in resolution and contrast with an improvement of 6 dB in CNR using only zero-angle transmission over the baseline. The observed reduction in the requirement of number of angles to produce similar image metrics can provide a possibility for higher frame rates.

*Clinical Relevance*— The proposed approach is an attempt towards realizing data adaptive beamforming algorithms such as MVDR at high frame rates employing CNNs. If achieved, could replace the conventional DAS based beamforming algorithms and offer better resolution and contrast images for clinical diagnostics.

## I. INTRODUCTION

Ultrasonography is one of the most popular diagnostic techniques in medical imaging due to its non-ionizing, non-invasive capability to acquire real time information on the state of the internal organs. Apart from the clinical research on adaptability to diagnose various aspects of pathology or lesions, this imaging modality has been an active research area in the domain of signal processing with a view to improve the quality of images acquired such that anatomic structures and anomalies are clearly delineated. Beamforming is an important step in image reconstruction for which different techniques like the DAS; which is the industry standard and its modified versions, e.g. delay multiply and sum (DMAS) [1] or MVDR; an adaptive beamformer are conventionally used [2]. In spite of the performance improvements offered by DMAS and MVDR beamformers, a much inferior DAS has been the approach employed in most of the clinical systems due to its low computational complexity and real-time performance. However, with the advent of learning-based approaches, recent studies have reported the use of neural networks (NNs) to accelerate image reconstruction process in US imaging [3-6]. A few examples of this are, in [3] where NNs are used in beamforming to minimize speckle noise, in [4] a learning-based model is used to make an adaptive beamformer where the apodization weights of pixel input data from the transducer are controlled by a FCNN and in [6] an attempt has been made to replace the conventional beamforming processes by an end-to-end deep learning neural network.

In this work, further improvisation has been carried out on the study in [4] using a region adaptive CNN architecture. The novelty of the proposed approach is with respect to the use of spatial information during the beamforming process which results in a two-fold advantage viz. (a) an equivalent image quality metric can be achieved with a lesser number of angles of transmission and thereby (b) improves the frame rate. The increase in frame rate will be critical for extension of learning algorithms in Doppler studies. The proposed approach combines the knowledge of MVDR beamforming and the efficiency of CNNs in order to produce fast and high-quality images. The paper is organized as follows. In Section II, the details of the proposed approach are presented. Section III describes the details of the experimental setup and Section IV discusses the results and the conclusions are presented in Section V.

## II. PROPOSED APPROACH

### A. CNN Architecture

The architecture of the CNN beamformer as shown in Fig. 1 consists of an input layer which is $L_2$ normalized with dimension $(x,y,c_h)$, where $x, y$ and $c_h$ are the axial and lateral coordinates and number of transducers (channels) respectively. Input normalization is done to bring stability to the training. For this work, the values for $(x,y,c_h)$ used are $(374,128,128)$. The input layer is followed by 4 hidden convolutional layers. The output layer of the network provides the adaptive apodization weights for each pixel in the input and hence the output dimension is the same as the input $(x,y,c_h)$.

The hidden layers are composed of 2D convolutional layers with number of filters increasing in dimension from 32 to 128 in factors of 2 at two steps. Each of the convolution layer has a filter kernel size of 3x3 and the output of the convolution i.e., the feature maps have the same height and width as that of the input. The activation function for the hidden layers is an anti-rectifier unit to prevent data being mapped to zero by the rectified linear unit (ReLU) for negative values that could appear in the hidden layers preventing dying nodes [4]. Batch normalization has been employed for each of the convolutional layers to further increase stability during training. The sum of apodization weights of all channels at the output layer for every pixel data is ensured to be unity with the application of softmax activation at the output layer in line with the unity constraint imposed in the MVDR. Due to the inherent design characteristics of the softmax function, which pushes higher values towards unity and lower values further towards zero, the contribution of the channel corrupted by noise towards the beamforming process gets automatically suppressed. At the same time, an inherent disadvantage is that it could lead to a situation where channels with a lower weight would be mapped to values that are closer to zero and channels which have a higher weight be mapped closer to unity. This behavior could have an impact on the resolution and contrast of the reconstructed image.

*Research supported by the Department of Science and Technology – Science and Engineering Research Board (DST-SERB (ECR/2018/001746)). Roshan P Mathews and Mahesh Raveendranatha Panicker are with Electrical Engineering and Center for Computational Imaging at the Indian Institute of Technology, Palakkad, Kerala, India. (corresponding author e-mail: mahesh@ iitpkd.ac.in).

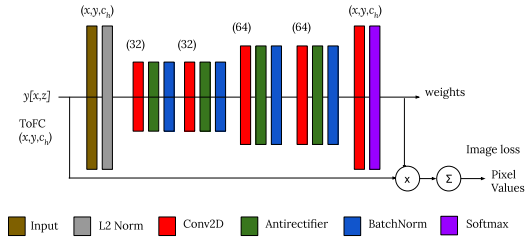

Figure 1. Proposed CNN architecture.

It has to be noted that the input to the CNN beamformer is the time of flight corrected (ToFC) data and the output of the CNN beamformer is the beamformed image prior to Hilbert Transform ie. the product of the output layer - the apodization weights and the input layer - the ToFC data are summed along the channel axis to obtain the beamformed image data. Upon this, the Hilbert transform is performed to obtain the final B-mode image.

*B. Comparison with the FCNN approach*

The architecture proposed in the baseline study is that of a FCNN [4] which does a pixel-by-pixel beamforming. The ToFC data is fed into a FCNN pixel by pixel (total $x*y$ pixels) with each pixel having contributions from $c_h$ number of channels. The motivation for the current work comes from the fact that in most imaging systems the value of a pixel is dependent on its neighborhood. Therefore, a CNN might be better suited to leverage the spatial information aspect during beamforming. CNNs with smaller kernel sizes e.g., 3x3 are ubiquitous and have been proven very successful in a wide variety of applications. Also, CNNs are translation equivariant and hence encode the same local features under translation creating an efficient encoding of the spatial information.

*C. Training Details*

The training input and output pair are the ToFC data and its corresponding MVDR beamformed image (prior to Hilbert transform). Thus, the error between the beamformed image of the proposed CNN beamformer and the MVDR are minimized on a loss function (image loss in Fig. 1) and the weights are updated through back propagation in each batch. The batch size was set to 30 scans (total training data included 600 ultrasound scans) in which the scans are selected at random and the loss function used was the Mean Squared Error (MSE) function. Alternatively, even a logarithmic loss function would also work well given that US images have a high dynamic range. The Adam optimizer was used with a learning rate of 1e-4. The models were trained with a stop criterion such that the mean squared error for the networks on test data were in the order of 1e-5 with the maximum error for a pixel in the order of 1e-3.

*D. Evaluation Details*

A detailed comparative evaluation of the outcome of the FCNN and CNN techniques in terms of the Contrast to Noise Ratio (CNR) and Full Width Half Maximum (FWHM) for observing the axial and lateral resolution has been done. The PICMUS data [7] acquired employing a plane wave transmission with 75 insonification angles [-16°,+16°] have been employed for the evaluation purpose. From the given 75 insonification angles, 3 cases were selected as follows for the image reconstruction to study the effect of number of transmission angles on the image metrics. The motivation here is to observe the CNN leverage spatial information and hence require lesser number of transmit angles to produce an image with superior quality.

- Case 1: Using only the 0° transmission – (single angle)
- Case 2: Using three angles of transmission (-0.43°, 0°, +0.43°) – (three angles compounded)

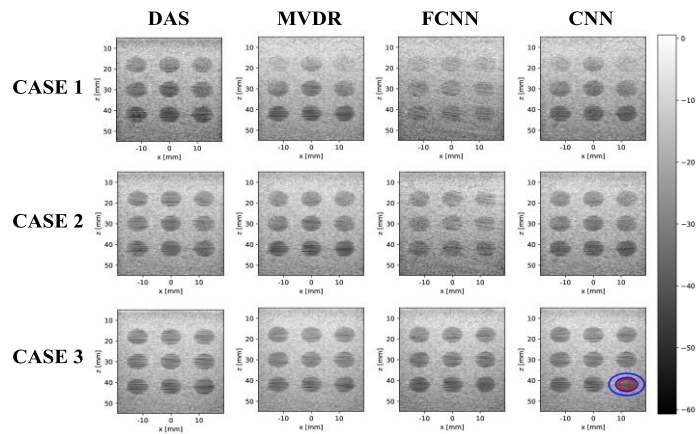

Figure 2. Comparison of raw beamformed images for Cases 1, 2 and 3 (different selection of insonification angles for beamforming) illustrated for the different beamformers discussed above. The ROI for the study is depicted in Case 3 CNN.

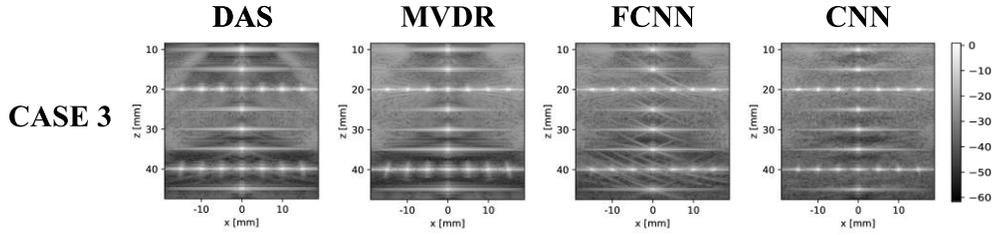

Figure 3. Results on the resolution and distortion PICMUS dataset for the beamformers in Case 3 (all insonification angles).

- Case 3: Using all seventy-five angles of transmission – (seventy-five angles compounded).

The PICMUS data which is sampled at a different rate is first resampled to match the rate of the training data (the sampling rate of the US machine used to acquire the training data) which was used to train the model. This inadvertently led to reconstruction artifacts like the extension of the dimension along the depth due to increasing samples which was removed by slicing the image data appropriately. For all the above cases the images are beamformed using the DAS, MVDR, FCNN and CNN on the same data.

### III. EXPERIMENTAL STUDY

The objective of this research work was to further improvise on the concept presented in [4] keeping it as a benchmark. The experimental study has been conceived in three parts. First, it required to confirm the implementation of the benchmark concept using an independent dataset collected during the current study. Second, an improvisation on the FCNN approach using a new CNN architecture has been proposed. Third, a comparative evaluation of the outcome of these techniques has been made in terms of qualitative and quantitative aspects. To maintain a common pivot while comparing the modalities, the same set of training and test data have been used for both the networks.

For the work reported herein, the data for training is obtained from the Verasonics research ultrasound machine (Verasonics Vantage 128 and L11-5v transducer, Verasonics, Kirkland, WA, USA) with center frequency of 7.6 MHz and sampling frequency of 31.25 MHz. Data was collected from 6 independent scans of 100 random frames each of the arm and finger is acquired from a healthy volunteer by following the principles outlined in the Helsinki Declaration of 1975, as revised in 2000. For test purposes, PICMUS dataset has been used. The choice of training and test data are such that the training and test set have no overlap (the source of the data generated for the training and test are different).

### IV. RESULTS AND DISCUSSION

The important results of the study are discussed here. Though images of all the case studies of all combinations of the ablation experiments are available, for the sake of brevity, only the essential images are presented. The resulting images for the DAS, MVDR, FCNN based beamformer [4] and the proposed CNN based beamformer are presented in Fig. 2 and 3. It has to be noted that herein, unlike traditional ultrasound images that employ image processing like speckle reduction or histogram corrections (thresholding, non-linear gamma correction and other similar techniques for image enhancement), the raw beamformed image is presented as it is to highlight the intrinsic differences among the beamformers.

From the results presented, it can be seen that the CNN based beamforming strategy clearly outperforms the FCNN method. This could be due to the fact that CNNs through their convolutional kernels, take into account the spatial coherence information when compared to FCNNs. For the Cases 1 and 2 shown in Fig. 2, the CNN reconstructed images are able to display the anechoic lesions better with lesser reconstruction artifacts while producing a good contrast when compared to the DAS for the particular region of interest (ROI) indicated in Fig. 2. It can also be seen that the CNN requires lesser number of transmit angles to produce a visually appealing image than the FCNN implementation. In fact, from our observations, the CNN is able to provide an equivalent CNR metric that is obtained with the FCNN which employs thrice as much insonification angles. With using just about one-fifth of the total insonification angles possible, the CNN provides superior image metrics to both the FCNN as well as the DAS too. This could imply that the proposed CNN approach can be employed for increased frame rate acquisition (lesser transmit angles implies the possibility of higher frame rate) when compared to the FCNN based approach. Although, with further increase in number of angles compounded, all modalities show improved sidelobes and contrast characteristics which is the expected behavior as seen in Fig. 3.

Given that visual inspection alone (qualitative measurement) cannot be the sole criteria to judge the quality of an image, the standard quantitative metrics such as the full width half maximum (FWHM) and contrast to noise ratio (CNR) are employed to judge the quality of the image as shown in Table 1. which reinforces our arguments. From the metric evaluation it is observed that the overall trend in this study is in agreement with the baseline study which serves as a validation for it and the CNN follows expected behavior of the minimum variance beamformer which it is trained on as evident from the superior lateral resolution which is a characteristic of the minimum variance beamformer. It is also observed that the study conducted on improvising the adaptive deep learning accelerated beamformer using a CNN produces metrics which display a significant improvement compared to the baseline study in terms of contrast and resolution for the cases selected here. The significant improvement in metrics, especially the CNR in cases 1 and 2 over the baseline study show promise to pursue more work in this area. Also, as a proof of demonstration for the completeness of the evaluation in an in-vivo setting, the carotid longitudinal section of the PICMUS dataset is presented in Fig. 4 which demonstrates superior quality on par with traditional beamformers in an in-vivo setting.

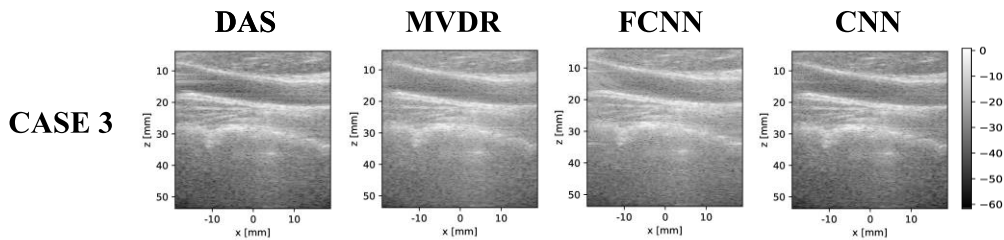

Figure 4. Results on the carotid longitudinal section of the PICMUS dataset for the completeness of the evaluation in an in-vivo setting.

TABLE I.     METRICS FOR DIFFERENT BEAMFORMERS

| Case # | Beamformer | FWHM (mm) Axial | FWHM (mm) Lateral | CNR (dB) |
|---|---|---|---|---|
| Case 1 | DAS | 0.388 | 0.946 | 6.882 |
| Case 1 | MVDR | 0.378 | 0.966 | 9.105 |
| Case 1 | FCNN | 0.407 | 1.001 | 2.555 |
| Case 1 | CNN | 0.417 | 0.926 | 8.418 |
| Case 2 | DAS | 0.404 | 0.966 | 7.037 |
| Case 2 | MVDR | 0.397 | 0.956 | 8.925 |
| Case 2 | FCNN | 0.417 | 1.005 | 4.431 |
| Case 2 | CNN | 0.400 | 0.926 | 8.844 |
| Case 3 | DAS | 0.455 | 0.887 | 8.112 |
| Case 3 | MVDR | 0.352 | 0.906 | 8.522 |
| Case 3 | FCNN | 0.401 | 0.966 | 9.536 |
| Case 3 | CNN | 0.420 | 0.852 | 8.624 |

On the computational complexity front, the CNN produces the complete apodization weights for an image in a single pass when compared to the FCNN and other pixel by pixel beamforming approaches that require multiple forward passes. In the CNN for a single pixel, the computational complexity is dominated by the convolution operations. The MACs (Multiply and Accumulate) for activations are lower order and hence while calculating the time complexity we can ignore them without causing drastic changes to the estimation of FLOPs (Floating Point Operations Per second). Going by this argument, the FLOPs for a single pixel in the reconstruction can be estimated as $\Sigma(K*K*C_i*C_{i+1})$, where: $K$ is the convolution filter kernel size, $C_i$ is the $i^{th}$ feature map filter dimension and $i$ goes from 0 (input) to 4 (output layer). This estimates to 313,344 FLOPs which is significantly lesser when compared to the regularized MVDR beamformer where the time complexity is dominated by the inversion of the covariance matrix ($O(n^3)$) which requires $128^3$ FLOPs for a single pixel reconstruction. This overall presents a three-fold advantage viz. (a) a high-quality image can be reconstructed in lesser time, (b) an equivalent image quality metric can be achieved with lesser number of transmit angles and thereby (c) improving the frame rate of acquisition.

## V. CONCLUSION

This paper demonstrates the usefulness of an improvised CNN based beamforming method for ultrasound image quality enhancements in terms of contrast and resolution. It is shown that the adaptive deep learning accelerated beamformer using a CNN outperforms the FCNN method in qualitative and quantitative sense for the cases selected here. Through this work a proof of concept for using region-based learning algorithms like the CNN beamformer is established. CNN beamformer is found to be capable of recreating a good contrast image in comparison with the conventional DAS. Another advantage noticed is that the region-based beamforming implementation requires a lesser number of angles and shows prospects of increased frame rate producing a better image in comparison with the FCNN and DAS. These findings probably open up further research in the area of automated ultrasound applications, Doppler, echocardiography and 3D reconstructions.

## ACKNOWLEDGMENT

Authors would like to acknowledge the funding from the Department of Science and Technology – Science and Engineering Research Board (DST-SERB (ECR/2018/001746)). The technical discussions with Mr. Ben Luijten and Ms. Gayathri Malamal are highly acknowledged. The implementation of the network is made open-sourced and available publicly [https://github.com/rpm1412/cnnusbf].